\documentclass[pdflatex,sn-mathphys-num]{sn-jnl}% Math and Physical Sciences Numbered Reference Style
\usepackage{graphicx}%
\usepackage{multirow}%
\usepackage{amsmath,amssymb,amsfonts}%
\usepackage{amsthm}%
\usepackage{mathrsfs}%
\usepackage[title]{appendix}%
\usepackage{xcolor}%
\usepackage{textcomp}%
\usepackage{manyfoot}%
\usepackage{booktabs}%
\usepackage{algorithm}%
\usepackage{algorithmicx}%
\usepackage{algpseudocode}%
\usepackage{listings}%
\usepackage{tikz}
\usetikzlibrary{decorations.pathreplacing}
\usepackage{array}

\theoremstyle{thmstyleone}%
\newtheorem{theorem}{Theorem}%  meant for continuous numbers
\newtheorem{proposition}[theorem]{Proposition}% 

\theoremstyle{thmstyletwo}%
\newtheorem{remark}{Remark}%

\theoremstyle{thmstylethree}%

\begin{document}
% \sethlcolor{yellow}

\title[Article Title]{General Quantum Protocol for Communication via $n$--Qubit GHZ States}

%%=============================================================%%
%% GivenName	-> \fnm{Joergen W.}
%% Particle	-> \spfx{van der} -> surname prefix
%% FamilyName	-> \sur{Ploeg}
%% Suffix	-> \sfx{IV}
%% \author*[1,2]{\fnm{Joergen W.} \spfx{van der} \sur{Ploeg} 
%%  \sfx{IV}}\email{iauthor@gmail.com}
%%=============================================================%%

\author*[1]{\fnm{Saba Arife} \sur{Bozpolat}}\email{saba.bozpolat@marmara.edu.tr}

\affil*[1]{\orgdiv{Physics Department}, \orgname{Marmara University}, \orgaddress{\street{Kuyubasi},
\postcode{34722}, \state{Istanbul}, \country{Turkiye}}}

\abstract{
This study presents a generalized $n$--bit superdense coding protocol that enables the transmission of n classical bits of information using an entangled $n$--qubit quantum system and the transmission of $n-1$ qubits. The protocol involves creating a maximally entangled $n$--qubit GHZ state, encoding the classical message with Pauli--Z and Pauli--X gates, and then transmitting and decoding the message via quantum communication, quantum operations, and measurements. The key novelty of this work is an explicit, scalable $n$--bit encoding routine, realized on the GHZ channel with a low--depth encoding layer using only single--qubit Pauli--X and Pauli--Z operations.

The protocol was tested on real quantum hardware using Qiskit 2.0 and the IBM--Fez quantum computer for message lengths of 4, 6, 8, and 10 bits, as a circuit--level proof of concept on a single processor; the physical transmission of qubits between the parties was not implemented. Results show that success rates decrease as message length, transpiled circuit depth and gate count increase; the latter two are set mainly by transpilation rather than by the encoding itself. A mutual information analysis shows that the transmitted information exceeds the classical bound of $n-1$ bits only for short messages. Strategies to improve performance include sending messages in shorter segments and advances in qubit coherence and gate fidelity. This work offers a practical and easily scalable quantum communication protocol with potential applications in quantum networks and communication systems.
}


\maketitle

\section{Introduction}\label{sec:intro}

The quantum superdense coding protocol \cite{bennett_1992} is a crucial method in quantum communication that significantly enhances channel capacity, as demonstrated experimentally with different systems such as photonic \cite{mattle_1996, schuck_2006, barreiro_2008, williams_2017, hu_2018, li_2025}, atomic \cite{schaetz_2004} and nuclear systems \cite{fang_2000}. In classical communication with a two--dimensional system, transmitting a single particle is limited to a maximum capacity of 1 bit. In contrast, within the quantum framework, if the two communicating parties, the sender Ay\c{c}a and the receiver Bora, share an entangled Bell state beforehand, Ay\c{c}a can perform local operations on their qubit to create four orthogonal Bell states and then send their qubit to Bora. Upon receiving the qubit, Bora performs a full Bell state measurement (BSM) to distinguish between the four states. This quantum strategy allows the transmission of 2 bits of information by sending only one particle, effectively doubling the channel capacity compared to classical limits \cite{bennett_1992, nielsen_2010}. In this respect, the protocol is not merely a theoretical concept but has significant application potential in the design of quantum networks, quantum cryptography, and quantum communication systems.

Recent research in the field of superdense coding is exploring its generalization to two--way and multi--particle communications, aiming to increase data rates and resource utilization even under realistic noise conditions, as well as expanding its capabilities by leveraging high--dimensional quantum systems \cite{hu_2018, jensen_2025, saha_2012, liu_2013, li_2025}. Both high--dimensional entanglement and multi--particle communication scenarios remain key enablers for surpassing previous channel capacity records and expanding practical quantum communication applications. While these developments are pushing us to integrate superdense coding into future quantum networks and communication systems, research to date does not provide an explicit, hardware--validated, gate--level construction for $n$--bit superdense coding on a single $n$--qubit GHZ channel.

The suitability of GHZ states as the entanglement resource is not self--evident. Liu et al.~\cite{liu_2013} establish that a pure state achieves optimal superdense coding, in the sense of 2 bits per qubit, if and only if the subsystem on which the encoding acts is maximally mixed, i.e. its von Neumann entropy equals its number of qubits ($n/2$ bits for an $n$--qubit state), a condition that GHZ states do not satisfy. However, this criterion pertains to a different communication scenario than the one considered here. The single--receiver dense coding capacity formula of Bru\ss~et al. \cite{bruss_2004, bruss_2005} yields exactly n bits for an $n$--qubit GHZ state, confirming the theoretical validity of such a protocol. GHZ states also offer practical advantages relevant to near--term hardware: they can be prepared with one Hadamard gate and $n-1$ CNOT gates, and their structure is directly compatible with the Pauli--X and Pauli--Z encoding routine introduced in this work. While alternative maximally entangled states, such as cluster \cite{raussendorf_2003} or $\chi$--type \cite{liu_2013} states, may satisfy different optimality criteria, this minimal preparation makes GHZ states particularly well suited to current NISQ hardware.

The remaining gap is constructive rather than theoretical. The capacity result of Bru\ss~et al. \cite{bruss_2004, bruss_2005} does not provide a gate-level construction. Saha and Panigrahi \cite{saha_2012} similarly present a mathematical encoding operator for a composite GHZ--Bell channel in closed form, but not an explicit circuit. Dutta et al. \cite{Dutta_2023}, motivated primarily by security rather than constructive realization, independently study the same N--qubit GHZ distribution and give an encoding rule using all three Pauli operators, with a four--way operator choice for the first qubit and a two--way choice for each remaining qubit, and with multiple valid operators permitted per message. Their treatment remains at the theoretical level, without an implementation or hardware validation. To our knowledge, previous hardware demonstrations of superdense coding on superconducting processors use Bell pairs \cite{zhukov_2019, garcia-martin_2018} or composite GHZ–Bell resources \cite{gautam_2019}; none implements $n$--bit dense coding on a single $n$--qubit GHZ channel. We close this gap on both fronts. Unlike the theoretical treatments above, our construction restricts itself to single--qubit Pauli--X and Pauli--Z operations under a single deterministic assignment that scales to any n, which keeps the resulting circuit within a low--cost, low--error native gate set on current NISQ hardware, and is matched to an explicit decoding circuit, $U^\dagger$ followed by a computational--basis measurement, from which Bora reads the message directly (Proposition~\ref{prop:correctness}). We further validate it experimentally on IBM--Fez for $n = 4, 6, 8, 10$ without physical qubit transmission, realizing, in the noiseless limit, the protocol that attains the Bru\ss~et al. theoretical capacity, and characterizing its degradation on a pure GHZ channel. The information--per--particle ratio of this protocol is $n/(n-1)$, which approaches 1 as $n$ grows. By the Holevo bound, $n-1$ particles transmitted without shared entanglement carry at most $n-1$ bits~\cite{nielsen_2010}. In the noiseless limit, the advantage of the protocol over this classical baseline is therefore exactly one bit for any $n$. By comparison, distributing $n/2$ Bell pairs and applying standard superdense coding to each pair transmits the same $n$ bits by sending only $n/2$ qubits instead of $n-1$~\cite{bennett_1992}, at the cost of the receiver storing $n/2$ entangled qubits rather than one. The ratio $n/(n-1)$ is not a deficiency but the defining feature of the single--receiver dense--coding regime of Bru\ss~et al. \cite{bruss_2004, bruss_2005}, whose capacity our construction realizes rather than exceeds. Our aim is constructive realizability; an explicit, hardware--executable gate--level construction; rather than resource optimality.

The flow of this paper is as follows: Section 2 describes the $n$--bit superdense coding protocol in stages. The first stage involves creating the maximally entangled $n$--qubit state. The second stage encodes the $n$--bit message into $n-1$ qubits using Pauli--Z and Pauli--X gates. The third and fourth stages involve transmitting the qubits and disentangling the quantum state to extract the message by making GHZ state measurements. The original idea of this work is the $n$--bit encoding routine described in the second stage. Section 3, then, presents test results on IBM--Fez, a real quantum device. The final section includes discussions and comments on the current results and directions for future work.

\section{Method}\label{sec:method}

This section describes the quantum algorithm for the $n$--bit superdense coding protocol in four stages. The first stage involves creating a maximally entangled $n$--qubit state in the quantum circuit. This stage is followed by the original part of this paper: the second stage describes how to encode an $n$--bit classical message using $n-1$ qubits. Then, the third stage requires qubit transmission via a suitable quantum communication channel. Finally, once all qubits have been received, Bora, the receiver, disentangles the qubits, performs GHZ State Measurements, and thus obtains the classical message.

We use the Qiskit (little--endian) ordering for which a computational basis state is written $|b\rangle = |b_{n-1}\cdots b_1 b_0\rangle$, where qubit $q_k$ carries bit $b_k$. Qubits $q_0,\dots,q_{n-2}$ belong to Ay\c{c}a and $q_{n-1}$ to Bora. For a bitstring $y$, $\bar{y}$ denotes its bitwise complement, and $\oplus$ is addition modulo 2.

\subsection*{Stage 1: Creating Quantum Entanglement}\label{sec:stage_1}

The first stage in the $n$--bit superdense coding protocol is to create a maximally entangled $n$--qubit state~\cite{bennett_1992,plenio_2007}. A multipartite entangled state can in principle serve as the entanglement resource, provided it is maximally entangled across the bipartition between the sender's and the receiver's qubits~\cite{bruss_2004,Dutta_2023}. In this work, we use the $n$--qubit GHZ state, which satisfies this condition, prepared by the circuit
\begin{equation}
  U \;=\; \Bigl(\prod_{k=1}^{n-1} \mathrm{CX}_{0\to k}\Bigr)\, H_0 ,
  \label{eqn:U}
\end{equation}
i.e.\ a Hadamard gate on qubit $q_0$ followed by CNOT gates from $q_0$ to each of the other qubits. Acting on $|0\rangle^{\otimes n}$, it yields the resource state
\begin{equation}
  |G_{0}\rangle \;\equiv\; U|0\rangle^{\otimes n}
  = \tfrac{1}{\sqrt{2}}\bigl(|0\rangle^{\otimes n}+|1\rangle^{\otimes n}\bigr).
  \label{eqn:G0}
\end{equation}
More generally, for an arbitrary input $|x\rangle=|x'\rangle|x_0\rangle$ with $x'=x_{n-1}\cdots x_1$,
\begin{equation}
  |G_x\rangle \;\equiv\; U|x\rangle
  \;=\; \tfrac{1}{\sqrt{2}}\Bigl(|x'\rangle|0\rangle + (-1)^{x_0}\,|\bar{x}'\rangle|1\rangle\Bigr).
  \label{eqn:GHZbasis}
\end{equation}
Since $U$ is unitary, the $2^n$ states $\{|G_x\rangle : x\in\{0,1\}^n\}$ form an orthonormal basis of the $n$--qubit Hilbert space, known as the GHZ basis, of which the resource state $|G_0\rangle$ is the $x=0\cdots0$ element. These are the states that Bora distinguishes in Stage 4, where $U^\dagger$ maps each of them back to the computational basis state with the same label:
\begin{equation}
  U^\dagger|G_x\rangle = |x\rangle .
  \label{eqn:inverseGHZbasis}
\end{equation}

We develop the explicit encoding routine (Stage 2) for the resource state $|G_0\rangle$. Extending the routine to other maximally entangled states, such as cluster or $\chi$--type states, may require modifying the encoding recipe and is left for future work.

For the superdense coding protocol to work, the entangled qubits created in the first stage must be shared between the two parties. To this end, the qubits may be entangled in a quantum hub and then distributed so that Ay\c{c}a, the sender, holds qubits $q_0,\dots,q_{n-2}$ and Bora, the receiver, holds qubit $q_{n-1}$. Since Bora's qubit remains untouched during encoding, the encoding routine in the next stage operates only on Ay\c{c}a's $(n-1)$--qubit register.

\subsection*{Stage 2: Encoding of Classical Message}\label{sec:stage_2}

\begin{figure}[t]
    \centering
    \begin{tikzpicture}
        \foreach \y [count=\i from 1] in {0,0.5} {
            \draw (0,\y) -- (8,\y);
            \node[left] at (0,\y) {$q_{n-\i}$};
        }
        \foreach \y [count=\i from 0] in {2.5,2.0,...,1.5} {
            \draw (0,\y) -- (8,\y);
            \node[left] at (0,\y) {$q_{\i}$};
        }
        \draw[double, double distance=1pt] (-0.5,3.20) -- (8,3.20);
        \draw[] (-0.35,3.1) -- node[above, font=\footnotesize] {$n$} (-0.15,3.3);
        \foreach \x [count=\i from 0] in {0.5,2.0,...,7.5} {
            \ifnum\i=3 \else
                \pgfmathsetmacro{\yf}{2.5-0.5*\i}
                \draw (\x,3.2) -- (\x,\yf);
                \ifnum\i=4
                    \fill (\x,3.2) circle (2pt) node[scale=0.5, below=14pt, right] {$n-2$};
                \else
                    \fill (\x,3.2) circle (2pt) node[scale=0.5, below=14pt, right] {$\i$};
                \fi
                \ifnum\i=0
                    \node[draw, fill=white, minimum width=0.3cm, minimum height=0.3cm] at (\x,\yf) {$Z$};
                \else
                    \node[draw, fill=white, minimum width=0.3cm, minimum height=0.3cm] at (\x,\yf) {$X$};
                \fi 
                \pgfmathsetmacro{\xp}{\x+0.7}
                \draw (\xp,3.2) -- (\xp,\yf);
                \fill (\xp,3.2) circle (2pt) node[scale=0.5, below=14pt, right] {$n-1$};
                \node[draw, fill=white, minimum width=0.3cm, minimum height=0.3cm] at (\xp,\yf) {$X$};
            \fi
        }
        \node at (0.0,1.1) {$\vdots$};    % dikey
        \node at (5.3,1.1) {$\ddots$};    % çapraz (sol üst → sağ alt)
        \draw[decorate, decoration={brace, amplitude=6pt}] (-1.0,0.4) -- (-1.0,3.4) node[midway, left=8pt] {Ay\c{c}a};
        \draw[decorate, decoration={brace, amplitude=6pt}] (-1.0,-0.15) -- (-1.0,0.3) node[midway, left=8pt] {Bora};
    \end{tikzpicture}
    \caption{Classical message encoding routine for the $n$--bit Superdense Coding Protocol. This routine constitutes the original proposal of this work and represents the second stage of the protocol.}
    \label{fig:fig1}
\end{figure}

This stage describes in detail how to implement the original method presented in this paper. Figure~\ref{fig:fig1} schematically illustrates how the $n$--bit superdense coding protocol should be implemented in a quantum circuit. As explained in Algorithm~\ref{alg:n_bit_superdense} explicitly, this protocol encodes an $n$--bit classical message by adding at most $2(n-1)$ single--qubit quantum gates to $n-1$ qubits, depending on the content of the message. Determined by the bit values in the message, the total number of quantum gates and the circuit depth will vary after the circuit is transpiled. In the best case scenario, all of the bits in the classical message are ``0''. For this case, the total number of gates added to the circuit during the encoding stage is zero, and the circuit depth is not affected. The worst case differs for gate count and for circuit depth. In terms of depth, after cancellation of repeated Pauli--X gates, the encoding adds at most two layers regardless of $n$, since only $q_0$ can receive two gates. This maximum is reached whenever the first and last bits are both ``1'', so that $q_0$ receives both a Pauli--Z and a Pauli--X, while every other qubit receives at most one Pauli--X. In terms of gate count, the same cancellation applies, and the relevant quantity is the count remaining after transpilation. This effective gate count is
\begin{equation}
G_{\mathrm{eff}}(b) = b_0 + b_{n-1} + \sum_{i=1}^{n-2}\big(b_i \oplus b_{n-1}\big),
\label{eqn:eff_gate_count}
\end{equation}
where the $b_{n-1}=1$ term inverts the contribution of the interior bits. Because of this cancellation the gate count is not monotonic in Hamming weight. The all--ones message, although it places the maximal $2(n-1)$ gates before transpilation, reduces to a single Pauli--Z and Pauli--X on $q_0$. The worst case is instead the message with ``1" in the first and last positions and ``0" elsewhere (e.g. 1001 for n=4), for which $q_0$ receives a Pauli--Z and a Pauli--X and each of the remaining $n-2$ qubits receives one Pauli--X, giving $n$ gates in total.

\begin{algorithm}[t]
\caption{Encoding Routine for the $n$--bit Superdense Coding Protocol}\label{algo1}
\begin{algorithmic}[1]
\Require quantum register with $n-1$ qubits, classical message of $n$--bit length.
\Ensure indexing starts with 0; the classical message is $b=b_{n-1}\cdots b_1 b_0$, and bit $b_k$ is assigned to qubit $q_k$.
\If {$b_0$ is ``1"}:
    \State add a Pauli--Z gate to $q_0$.
\EndIf
\If {$b_{n-1}$ is ``1"}:
    \State add a Pauli--X gate to $q_0$.
\EndIf	
\For {$i=1 \text{ to } n-2$}:
    \If {$b_i$ is ``1"}:
        \State add a Pauli--X gate to $q_i$.
    \EndIf
    \If {$b_{n-1}$ is ``1"}:
        \State add a Pauli--X gate to $q_i$.
    \EndIf
\EndFor
\end{algorithmic}
\label{alg:n_bit_superdense}
\end{algorithm}

\subsection*{Stage 3 \& 4: Quantum Communication and Disentangling}\label{sec:stage_3-4}

In the first one of the last two stages, Ay\c{c}a sends their $n-1$ qubits to Bora. This process requires the use of appropriate quantum communication channels. And finally at the last stage, the maximum entanglement in the state of the qubits in the circuit is resolved by reversing the first stage of the $n$--bit superdense coding protocol. This is achieved by adding the quantum operations from Stage--1 to the quantum circuit in mirror--symmetric order.

\subsection*{Correctness of the protocol}\label{sec:correctness}

For a message $b\in\{0,1\}^n$, Algorithm 1 applies, on Ay\c{c}a's qubits only,
\begin{equation}
  E_b \;=\; \Bigl(\prod_{i=1}^{n-2} X_i^{\,b_i\oplus b_{n-1}}\Bigr)\; X_0^{\,b_{n-1}}\, Z_0^{\,b_0}.
  \label{eqn:Eb}
\end{equation}
To follow the result of $X$ gates more easily, we define a flip pattern $f=f_{n-1}\cdots f_0$ with,
\begin{equation*}
  f_0=b_{n-1},\qquad f_i=b_i\oplus b_{n-1}\ (1\le i\le n-2),\qquad f_{n-1}=0.
\end{equation*}
With that, $E_b$ becomes,
\begin{equation}
  E_b \;=\; \Bigl(\prod_{i=0}^{n-2} X_i^{\,f_i}\Bigr)\, Z_0^{\,b_0}.
  \label{eqn:Eb}
\end{equation}

\begin{proposition}
For every $b\in\{0,1\}^n$, $E_b|G_0\rangle = (-1)^{b_0 b_{n-1}}\,|G_b\rangle$. Consequently, in the noiseless limit, Bora's measurement of $U^\dagger E_b |G_0\rangle$ in the computational basis returns $b$ with probability $1$, and the map $b\mapsto E_b|G_0\rangle$ perfectly distinguishes all $2^n$ messages.
\label{prop:correctness}
\end{proposition}

\begin{proof}
Applying $Z_0^{b_0}$ to $|G_0\rangle$ gives $\tfrac{1}{\sqrt{2}}\bigl(|0\rangle^{\otimes n}+(-1)^{b_0}|1\rangle^{\otimes n}\bigr)$. The $X$ gates in Eq.~\ref{eqn:Eb} then flip the qubits and the result is
\begin{equation}
  E_b|G_0\rangle=\tfrac{1}{\sqrt{2}}\bigl(|f\rangle+(-1)^{b_0}|\bar f\rangle\bigr).
  \label{eqn:afterE}
\end{equation}
We compare Eq.~\ref{eqn:afterE} with Eq.~\ref{eqn:GHZbasis}, whose first term is the component with $q_0$ in $|0\rangle$.

\emph{Case $b_{n-1}=0$.} Then $f_0=0$ and $f=0\,b_{n-2}\cdots b_1\,0$, i.e. $|f\rangle=|b'\rangle|0\rangle$ with $b'=b_{n-1}\cdots b_1$. Eq.~\ref{eqn:afterE} reads $\tfrac{1}{\sqrt{2}}\bigl(|b'\rangle|0\rangle+(-1)^{b_0}|\bar b'\rangle|1\rangle\bigr)=|G_b\rangle$.

\emph{Case $b_{n-1}=1$.} Then $f_0=1$, and the complement satisfies $\bar f_0=0$, $\bar f_i=b_i$ for $1\le i\le n-2$, $\bar f_{n-1}=1$, i.e. $|\bar f\rangle=|b'\rangle|0\rangle$. Factoring out $(-1)^{b_0}$,
\begin{equation*}
  E_b|G_0\rangle=(-1)^{b_0}\,\tfrac{1}{\sqrt{2}}\bigl(|b'\rangle|0\rangle+(-1)^{b_0}|\bar b'\rangle|1\rangle\bigr)
  =(-1)^{b_0}|G_b\rangle .
\end{equation*}
Both cases give $E_b|G_0\rangle=(-1)^{b_0b_{n-1}}|G_b\rangle$. Stage 4 applies $U^\dagger$, yielding $(-1)^{b_0b_{n-1}}|b\rangle$; the global phase is unobservable, so the computational--basis measurement returns $b$ with certainty. Since $b\mapsto|G_b\rangle$ is a bijection onto an orthonormal basis, distinct messages are mapped to orthogonal states and are therefore perfectly distinguishable.
\end{proof}

\begin{remark}
$E_b$ acts only on $q_0,\dots,q_{n-2}$, so Bora's qubit $q_{n-1}$ is never touched and $n$ classical bits are conveyed by transmitting $n-1$ qubits. After cancellation of repeated $X$ gates, $E_b$ contains at most $n$ single--qubit Pauli gates, attained for $b_0=b_{n-1}=1$ and $b_i=0$ ($1\le i\le n-2$).
 
The result extends to any known GHZ--basis resource $|G_x\rangle$. Since $E_b=X^{f(b)}Z_0^{b_0}$ with $f(b)$ and $b_0$ linear in $b$ over $\mathbb{Z}_2$, and $Z_0X_0=-X_0Z_0$, one has $E_bE_x=\pm E_{b\oplus x}$; with $|G_x\rangle=\pm E_x|G_0\rangle$ from Proposition~\ref{prop:correctness},
\begin{equation}
  E_b|G_x\rangle=\pm|G_{b\oplus x}\rangle ,
  \label{eqn:EbGx}
\end{equation}
so Bora measures $b\oplus x$ with certainty and recovers $b$ by a bitwise XOR with $x$. This is relevant to Pauli--frame tracking on hardware, since a resource prepared as $|G_x\rangle$ because of a known Pauli correction or error need not be re--prepared.
\end{remark}

Ultimately, $n$--bits of information were encoded into qubits using an $n$--qubit GHZ system with this novel quantum superdense coding protocol. In this way, the protocol transmits $n$ bits of information from the sender (Ay\c{c}a) to the receiver (Bora) by sending $n-1$ qubits through a suitable quantum communication channel. The most critical aspect of the protocol described here is that it proposes a relatively simple and easily scalable method for quantum scenarios that require a bit encoding procedure.

\section{Results}\label{sec:results}

This section presents the results of quantum calculations run on a real quantum computer using different classical message lengths. We considered messages of 4, 6, 8, and 10 bits in length in these  calculations. We implemented the protocol in Qiskit 2.0 and executed it on the IBM--Fez quantum processor (Heron r2), whose specifications are listed in Table~\ref{tab:tab1}. The resulting dataset is available in \cite{zenodo_link}.

\begin{table}[t]
    \setlength{\aboverulesep}{0pt}
    \setlength{\belowrulesep}{0pt}
    \renewcommand{\arraystretch}{1.3}
    \newcommand{\cw}[1]{\dimexpr(\linewidth-0.7cm-10\tabcolsep)*#1/100\relax}
    \caption{Technical specification of IBM--Fez quantum processor. This quantum machine is used to run the quantum codes produced in this work. CLOPS is short for ``Circuit Level Operations per Second''.}
    \centering
    \begin{tabular}{p{\cw{20}}p{\cw{10}}p{\cw{20}}@{\hspace{0.7cm}}p{\cw{24}}p{\cw{6}}p{\cw{20}}}
    \cmidrule(r{0.7cm}){1-3}\cmidrule{4-6}
    Processor Type & \multicolumn{2}{l}{Heron r2} &
        \multicolumn{2}{@{}l}{Median Readout Error} & 1.45E-2 \\ \cmidrule(r{0.7cm}){1-3}\cmidrule{4-6}
    Total Qubits & \multicolumn{2}{l}{156} &
        \multicolumn{2}{@{}l}{Median T$_1$ ($\mu$s)} & 145.51 \\ \cmidrule(r{0.7cm}){1-3}\cmidrule{4-6}
    Total Couplers & \multicolumn{2}{l}{176} &
        \multicolumn{2}{@{}l}{Median T$_2$ ($\mu$s)} & 105.97 \\ \cmidrule(r{0.7cm}){1-3}\cmidrule{4-6}
    CLOPS & \multicolumn{2}{l}{320 K} &
        \multicolumn{2}{@{}l}{2Q Error (Best)} & 1.35E-3 \\ \cmidrule(r{0.7cm}){1-3}\cmidrule{4-6}
    \multirow{2}{*}{Basis Gates} & 1--qubit & id, rx, rz, sx, x &
        \multirow{2}{*}{$\begin{aligned}&\text{Median Errors for}\\&\text{Basis Gates}\end{aligned}$} & sx & 3.239E-4 \\ \cmidrule(r{0.7cm}){2-3}\cmidrule{5-6}
    & 2--qubit & cz, rzz &
        & cz & 2.585E-3 \\ \cmidrule(r{0.7cm}){1-3}\cmidrule{4-6}
    \end{tabular}
    \label{tab:tab1}
\end{table}
 
\begin{figure}[b]
\centering
\includegraphics[width=1.0\textwidth]{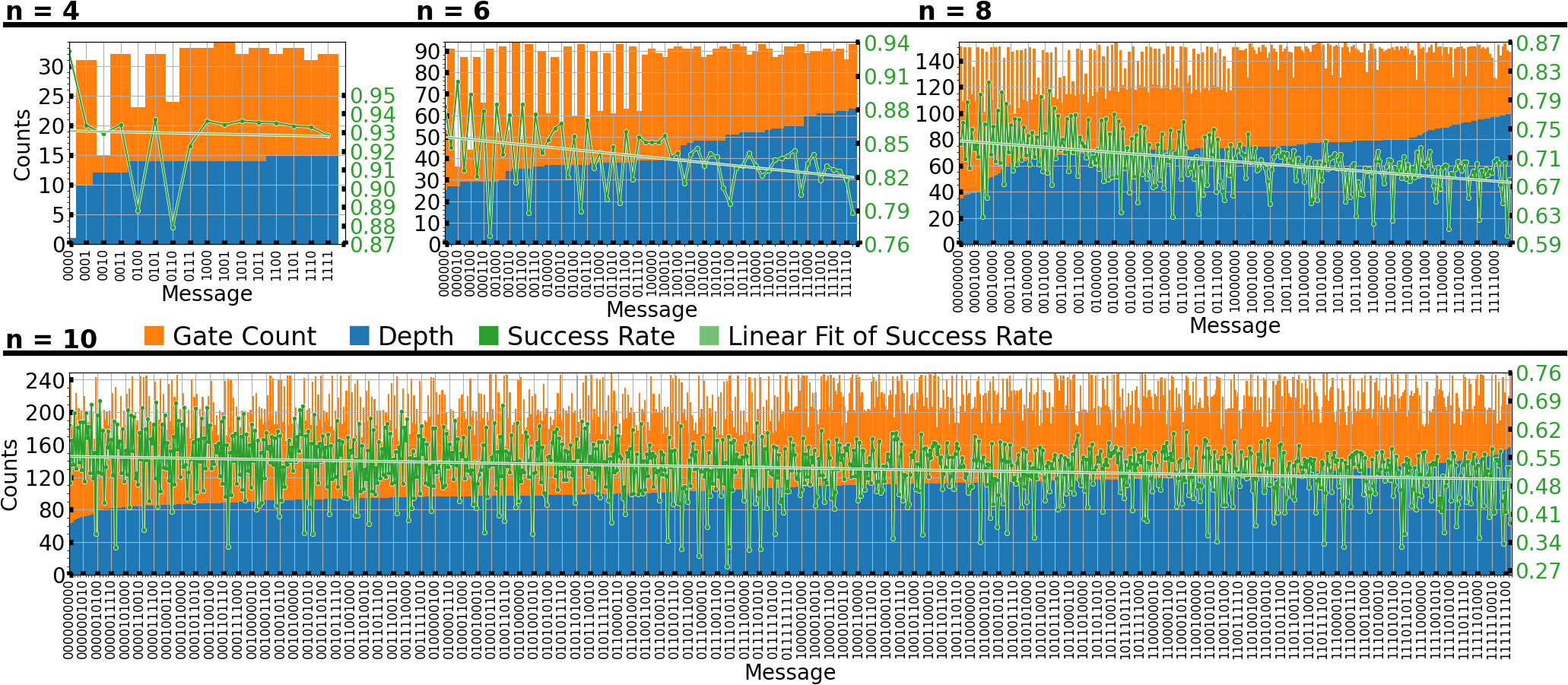}
\caption{(colored online) Results showing (orange) gate counts, (blue) depth and (green) success rate vs initial message bitstring for (a) 4--bit, (b) 6--bit, (c) 8--bit and (d) 10--bit in length.}
\label{fig:fig2}
\end{figure} 

All quantum circuits used in computations contain the same number of qubits as the length of the corresponding classical message. For a classical message containing $n$ bits of information, $2^n$ different bitstrings are possible. In this study, a separate quantum circuit containing $n$ qubits was created and the $n$--bit superdense coding protocol was implemented for each one of these bitstrings. In each quantum circuit, maximum entanglement was achieved using the entanglement routine explained in Stage--1. Then, the second stage of the $n$--bit superdense coding protocol was applied to encode the relevant classical message into the quantum state of the circuit. Because it requires physical quantum communication, Stage--3 is omitted. The final stage was applied to the quantum circuit in order to disentangle the statevector and then, measurements were added to all qubits to determine whether the initial classical message is obtained. Each quantum circuit was run 4096 times, and the resulting bitstrings were accumulated per quantum circuit. Based on these accumulations, the success rate of receiving the initial classical message was calculated. Then these success rates were compared across both circuit depths and gate counts. The reported success rates characterize the full circuit execution (entanglement, encoding, and disentangling) excluding only the physical transmission stage.

Figure~\ref{fig:fig2} shows that, for $n \geq 6$, the success rate tends to decrease as the decimal value of the transmitted message increases. This trend cannot be attributed to the encoding itself, whose effective gate count is not monotonic in the message value because repeated Pauli--X gates on the same qubit cancel (see Eq.~\ref{eqn:eff_gate_count}). Instead, it follows the transpiled circuit, as both the circuit depth and the total gate count after transpilation tend to increase with the decimal value of the message.

\begin{table}[t]
\caption{Messages taken into account in the analysis for Figure \ref{fig:fig3}. For the initial messages with 50\% Hamming weight, we have calculated the average of success rates of all possible initial messages. Check the main text for more detail.}
\begin{tabular}{clll}
\cmidrule[1.3pt]{1-4}
\multirow{2}{*}{Length} & \multicolumn{3}{l}{Hamming Weight Percentage} \\ \cmidrule{2-4}
 & \multicolumn{1}{l}{0\% } & \multicolumn{1}{l}{50\%} & 100\% \\ \cmidrule[1.3pt]{1-4}
4 & \multicolumn{1}{l}{``0000"} & \multicolumn{1}{l}{``0011", ``0101", ``0110", ``1001", ``1010", ``1100"} & ``1111" \\ \cmidrule{1-4}
6 & \multicolumn{1}{l}{``000000"} & \multicolumn{1}{l}{``000111", ``010101", ``101010", $\dots$} & ``111111" \\ \cmidrule{1-4}
8 & \multicolumn{1}{l}{``00000000"} & \multicolumn{1}{l}{``00001111", ``01010011", $\dots$} & ``11111111" \\ \cmidrule{1-4}
10 & \multicolumn{1}{l}{``0000000000"\;\;\;\;} & \multicolumn{1}{l}{``0000011111", ``0101010110", $\dots$ \;\;\;\;} &  ``1111111111"\;\;\;\; \\ \cmidrule[1.3pt]{1-4}
\end{tabular}
\label{tab:table2}
\end{table} 

\begin{figure}[b]
\centering
\includegraphics[scale=0.35]{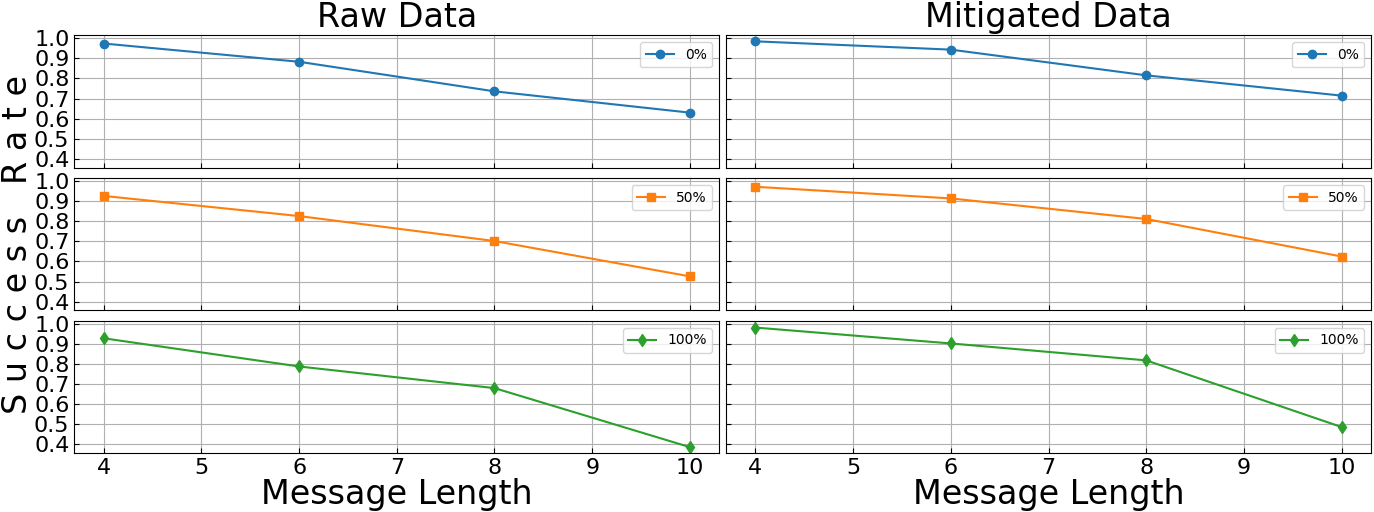}
\caption{(colored online) Success rate change with respect to the message length. Hamming weight percentages in the messages are (blue, circle) 0\%, (orange, square) 50\%, and (green, diamond) 100\%. Check the main text for more detail. }
\label{fig:fig3}
\end{figure}

\begin{figure}[t]
\centering
\includegraphics[width=\textwidth]{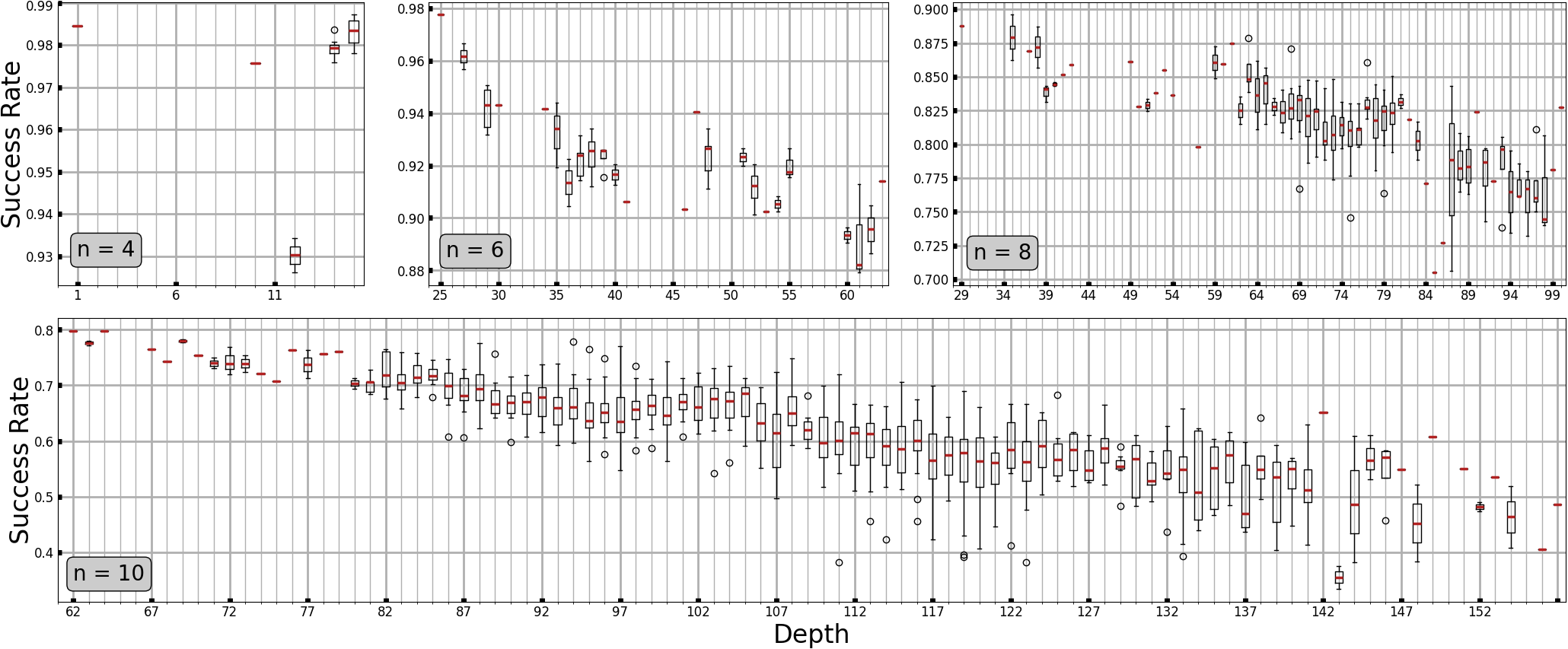}
\caption{(colored online) Results showing success rate vs depth for (a) 4--bit, (b) 6--bit, (c) 8--bit and (d) 10--bit length initial messages. Error mitigated data is used to generate this plot.}
\label{fig:fig4}
\end{figure}

\begin{figure}[t]
\centering
\includegraphics[width=\textwidth]{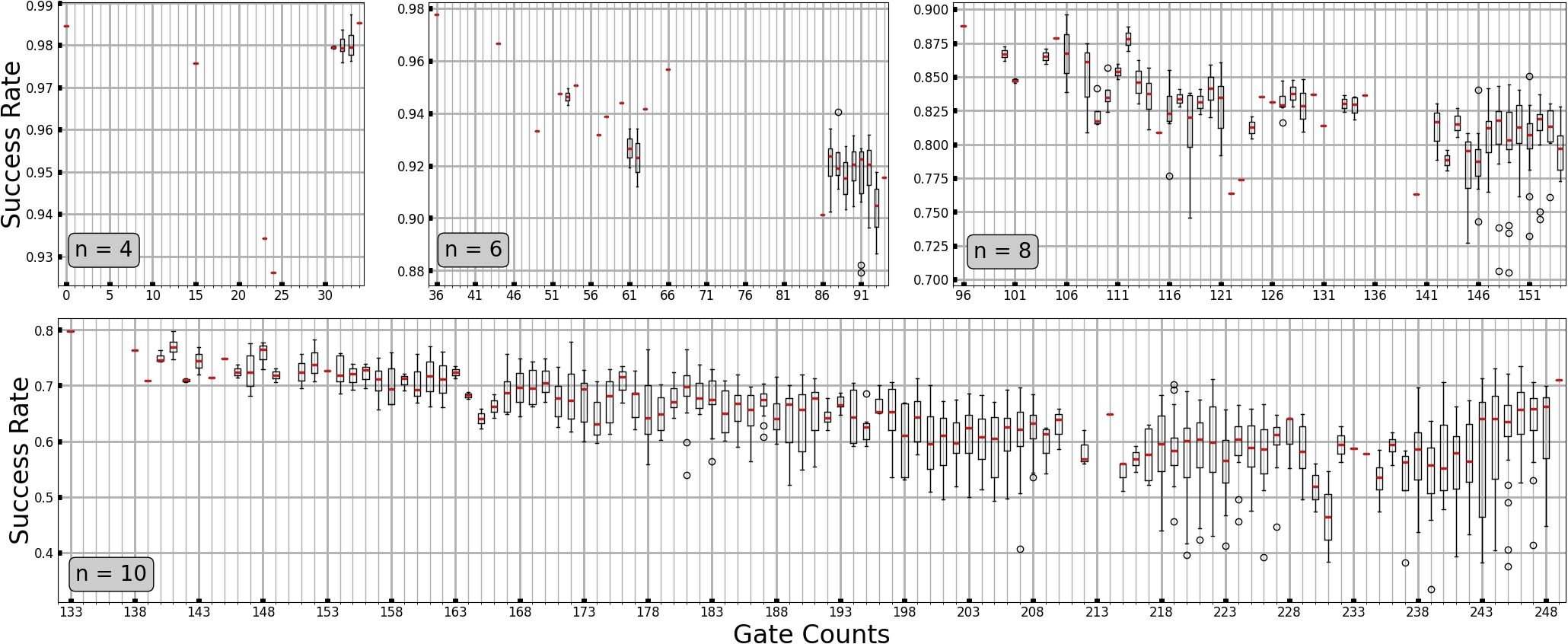}
\caption{(colored online) Results showing success rate vs gate count for (a) 4--bit, (b) 6--bit, (c) 8--bit and (d) 10--bit length initial messages. Error mitigated data is used to generate this plot.}
\label{fig:fig5}
\end{figure}

One approach for improving the success rate would be to produce qubits with longer coherence times and fewer errors. As more reliable qubits are produced, therefore the success rates of this protocol will naturally increase. Another effective strategy to improve the success rates is to transmit the desired message in shorter chunks rather than as a single long message. According to Figure~\ref{fig:fig2}, this approach demonstrably increases the transmission success rate, which becomes more apparent in Figure~\ref{fig:fig3}. In addition to these, the reported success rates can be improved by applying certain error mitigation techniques. In this work, we employed readout error mitigation as a straightforward implementation, and the resulting improvement in the outcomes is demonstrated in Figure~\ref{fig:fig3}. As expected, the application of readout error mitigation increased the probability of transmitting the correct message. The improvement from mitigation is partial rather than complete, and the qualitative degradation trend with n persists after mitigation, indicating that gate errors and decoherence accumulated across the full protocol run, rather than readout error alone, are the dominant noise contributors. For the $n=10$ all--ones message, which shows the lowest success rate in Figure 3, the mitigated success rate remains just below 0.5. The correct message is therefore no longer recovered in the majority of shots, yet it remains the single most frequent outcome (see Ref \cite{zenodo_link}) and stays far above the $2^{-n}$ random--guessing baseline ($2^{-10} \approx 0.001$ for n = 10). This underscores that longer messages approach the practical limits of current hardware. It should be noted that Figure~\ref{fig:fig2} is generated from raw data; the error--mitigated counterpart is provided in the supplementary material.

Figure~\ref{fig:fig3} shows the variation in success rates depending on message length. This graph directly examines the success rates of three different message types, namely those with 0\% (blue, circle), 50\% (orange, square), and 100\% (green, diamond) Hamming weight values. The 100\% and 0\% Hamming weight message types each correspond to a unique bitstring. However, for the message type with 50\% Hamming weight, multiple distinct bitstrings exist for each length. Accordingly, the success rates were averaged over all such bitstrings for each message length. Details of the messages used in this graph are found in Table~\ref{tab:table2}. Note the decrease in success rate as message length increases. This trend supports the idea that transmitting messages in shorter chunks leads to reduced information loss and improved transmission accuracy. However, the optimal chunk length is an important open question, as it directly affects both the protocol efficiency and the initial qubit sharing between the parties, which is beyond the scope of this study.

In addition to the aforementioned analyses, we investigated the dependence of the success rate on circuit depth and gate count. Since a given depth or gate count value generally corresponds to multiple classical messages with differing success rates, the success rates were aggregated accordingly, as presented in Figures~\ref{fig:fig4} and~\ref{fig:fig5}. Consistently with the preceding results, the success rate tends to decrease with increasing gate count and circuit depth.

\begin{figure}[t]
\centering
\includegraphics[width=0.6\textwidth]{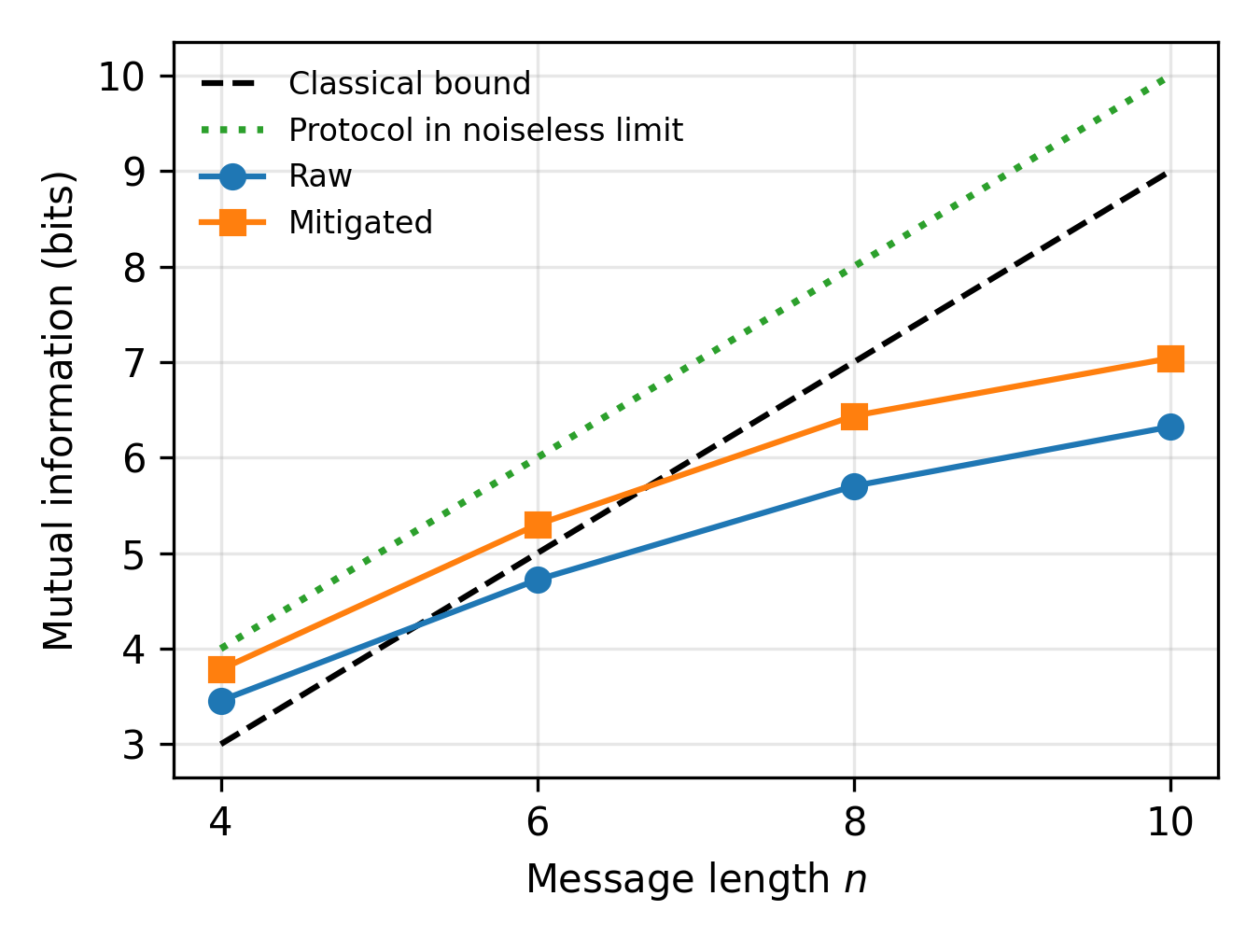}
\caption{(colored online) Mutual information between sent and received messages versus message length $n$, computed from raw (blue, circle) and readout--mitigated (orange, square) IBM--Fez data, assuming equiprobable messages. The dashed line marks the classical bound of $n-1$ bits for transmitting $n-1$ particles without shared entanglement; the dotted line marks the $n$ bits carried by the protocol in the noiseless limit.}
\label{fig:fig6}
\end{figure}

Finally, the success rates can be translated into transmitted information. A more stringent baseline than random guessing is the classical transmission of the same $n-1$ particles, which carries at most $n-1$ bits. Figure~\ref{fig:fig6} compares this bound with the mutual information between sent and received messages, estimated from the measured distributions. For each $n$, the conditional probabilities $p(\hat{b}|b)$ of measuring $\hat{b}$ when $b$ is sent are taken from the normalized counts of the $2^n$ circuits, and, for equiprobable messages,
\begin{equation}
  I(B;\hat{B}) = H(\hat{B}) - \frac{1}{2^n}\sum_{b} H(\hat{B}|B=b),
  \label{eq:mutual_info}
\end{equation}
where $H$ denotes the Shannon entropy in bits.

As shown in Figure~\ref{fig:fig6}, the shortfall from the $n$ bits attained in the noiseless limit grows with $n$. Since the classical bound lies exactly one bit below the noiseless limit, the protocol retains its advantage only while this shortfall remains below one bit. This holds for $n=4$ in both raw and mitigated data and for $n=6$ only after mitigation, whereas for $n \geq 8$ the protocol falls below the classical bound. Readout mitigation increases the mutual information for all $n$ but does not lift the $n \geq 8$ values above the bound, indicating that gate errors and decoherence, rather than readout error, are the dominant limitation. On current hardware, the one--bit quantum advantage is therefore confined to short messages. This supports the chunking strategy discussed above, since transmitting a long message in short blocks preserves the advantage that a single long block loses.

\section{Conclusion}\label{sec:conc}

The results demonstrate that the success rate of transmitting classical messages using the $n$--bit Superdense Coding Protocol on a quantum computer decreases as the message length, circuit depth, and gate count increase. Circuit depth and gate count are set mainly by transpilation rather than by the encoding itself, whose effective gate count is not monotonic in the message value (see Eq.~\ref{eqn:eff_gate_count}); longer transpiled circuits accumulate more gate errors and decoherence. Improving qubit coherence and fidelity remains a key avenue for enhancing protocol performance. This also implies a practical, hardware--set ceiling on the message length $n$ that can be reliably transmitted with current devices, rather than any limitation intrinsic to the encoding routine. The mutual information analysis makes this ceiling quantitative, since the protocol exceeds the classical bound of $n-1$ bits only for short messages ($n=4$; $n=6$ only after mitigation). Additionally, dividing longer messages into shorter chunks shows promise for increasing transmission accuracy, although determining the optimal chunk size is a subject for future research. Three more natural directions remain open. The first is a complete end--to--end hardware demonstration including the physical transmission stage, which would also be the natural setting for a systematic security analysis against joint or adaptive interception strategies; as with any superdense coding scheme, the protocol alone provides no security against eavesdropping, which requires an additional channel--verification step such as those of Refs.~\cite{bostrom_2002, deng_long_2004, Dutta_2023}. The second is the application of the same encoding routine to other maximally entangled states such as cluster or $\chi$--type states. A third is a full noise--channel characterization that separates individual error contributions, such as $T_1/T_2$ decay versus gate infidelity, to identify which improvements would most directly extend the practical ceiling on $n$.

Overall, this work provides an explicit, scalable gate--level construction of the $n$--bit superdense coding protocol. It closes the gap left by prior theoretical treatments of the single GHZ channel, which establish capacity or closed--form operators without an executable realization. Beyond this, the work highlights the trade--offs between circuit complexity and communication reliability on existing quantum hardware.

\section*{Acknowledgements}\label{sec:ack}

The research question addressed in this study arose from scientific discussions held within the scope of the Introduction to Quantum Computing undergrad course offered in the 2024--2025 spring semester at the Marmara University, Department of Physics. We would like to thank the students who took the course, Ayşe Esmanur Şafak, Beyza Kayalan, İlknur Derman, Çayan Kuzu, Kutluhan Tuncay, Muhammed Raşit Yurdakul, and Ömer Faruk Okal, for their motivation and contributions to the discussions throughout the semester.

\section*{Data availability}
The dataset created and analyzed within the scope of this study can be accessed at Ref \cite{zenodo_link}.

\section*{Funding}
No part of this work has received any financial support from any funding agency and/or other organization.

\bibliography{sn-article}% common bib file

\end{document}